\documentclass[a4paper,aps,prd,10pt,preprintnumbers,showpacs,twocolumn,superscriptaddress,nofootinbib,amsmath,amssymb]{revtex4-1}
\usepackage{graphicx}
\usepackage{epstopdf}
\usepackage{cmap}
\usepackage[utf8]{inputenc}
\usepackage[T1]{fontenc}
\def\imo{i}
\def\re#1{Re(#1)}
\def\im#1{Im(#1)}
\def\K{{\cal K}}
\def\Order#1{{\cal O}\left(#1\right)}

\begin{document}
\title{Long-lived massive scalar modes, grey-body factors, and absorption cross sections of the Reissner--Nordstr\"om-like brane-world black hole}
\author{Zainab Malik}\email{zainabmalik8115@outlook.com}
\affiliation{Institute of Applied Sciences and Intelligent Systems, H-15 Islamabad, Pakistan}
\begin{abstract}
We study quasinormal modes, including the quasi-resonant regime, grey-body factors, and absorption cross sections of a massive scalar field in a Reissner--Nordstr\"om-like brane-world black hole endowed with a tidal-charge parameter induced by extra-dimensional effects. Combining semiclassical WKB calculations with time-domain evolution, we determine the range of parameters for which the effective potential keeps the single-barrier shape needed for a reliable quasinormal-mode and scattering analysis. We find that increasing positive tidal charge lowers the barrier, drives the spectrum closer to the quasi-resonant regime, and enhances transmission and absorption, whereas increasing the field mass or multipole number makes the barrier less transparent and shifts absorption to higher frequencies. Our results indicate the onset of an arbitrarily long-lived quasinormal-mode regime. At the same time, this behavior cannot be followed directly in the time-domain profiles, because the asymptotic tails set in too early and mask the late-time ringing.
\end{abstract}
\maketitle
\section{Introduction}

Quasinormal modes are the characteristic damped oscillations that govern the relaxation of a black hole after it is perturbed. Their complex frequencies are determined by the background geometry, the nature of the perturbing field, and the physically motivated boundary conditions at the horizon and at infinity, which makes them a natural tool for probing dynamical stability, interpreting ringdown signals, and discriminating between different black-hole geometries and extensions of general relativity \cite{Kokkotas:1999bd,Berti:2009kk,Konoplya:2011qq,Bolokhov:2025rng}. For this reason the quasinormal spectrum has become one of the standard diagnostics in black-hole physics: any ingredient that modifies the effective potential or the asymptotic behavior of perturbations is expected to leave a clear imprint on the oscillation frequency and damping rate.

Quasinormal modes of massive fields furnish a particularly interesting case because spectrum of massive fields have distinctive behaviour both in the time and frequency domains. For that reason massive perturbations have been studied in a wide variety of settings, including charged and rotating black holes, higher-dimensional spacetimes, regular geometries, and magnetized configurations \cite{Konoplya:2004wg,Aragon:2020teq,Ponglertsakul:2020ufm,Konoplya:2018qov,Konoplya:2017tvu,Zhidenko:2006rs,Ohashi:2004wr,Gonzalez:2022upu}. 

There are several reasons why a massive scalar field is a natural object to study. In higher-dimensional and brane-based scenarios, effective mass terms may arise from the influence of bulk degrees of freedom on the four-dimensional brane dynamics \cite{Seahra:2004fg}. More generally, massive propagating sectors, including effective massive-graviton descriptions, are relevant to present-day gravitational-wave phenomenology and to the interpretation of very low-frequency signals such as those discussed in the context of pulsar timing observations \cite{Konoplya:2023fmh,NANOGrav:2023hvm}. Even when the fundamental perturbing field is massless, the surrounding environment can generate an effective mass term, as happens for black holes in external magnetic fields \cite{Konoplya:2008hj,Wu:2015fwa}. Thus the massive field's problem is not merely a toy model, but a flexible effective description of several physically motivated situations.

The background considered in this paper is also motivated by more than one physical picture. The metric function $f(r)=1-2M/r-\gamma/r^2$ is formally of Reissner--Nordstr\"om type, yet the coefficient of the $r^{-2}$ term need not represent an electric charge. In particular, Reissner--Nordstr\"om-like geometries emerge in the braneworld construction of Dadhich, Maartens, Papadopoulos, and Rezania, where the charge-like parameter is reinterpreted as a tidal charge induced by the bulk geometry rather than by a Maxwell field \cite{Dadhich:2000am,Seahra:2004fg}. This makes the RN-like background a useful one-parameter deformation of Schwarzschild for testing how charge-like corrections, whether electromagnetic or purely geometric, affect the quasinormal spectrum of a massive field.

One of the central features of massive perturbations is the possibility of quasi-resonances, that is, modes for which the damping rate becomes extremely small while the oscillation frequency remains finite. This behavior was identified early for massive scalar perturbations and later found for fields of different spin \cite{Ohashi:2004wr,Konoplya:2004wg,Fernandes:2021qvr,Konoplya:2017tvu,Percival:2020skc}. Subsequent studies showed that long-lived modes arise in a remarkably broad class of black-hole backgrounds and even for other compact objects such as wormholes \cite{Lutfuoglu:2025qkt,Bolokhov:2023ruj,Skvortsova:2025cah,Lutfuoglu:2026xlo,Zhidenko:2006rs,Bolokhov:2026dzn,Skvortsova:2026unq,Dubinsky:2025bvf,Bolokhov:2024bke,Lutfuoglu:2025hwh,Zinhailo:2018ska,Churilova:2020bql,Bolokhov:2023bwm,Lutfuoglu:2026fpx,Skvortsova:2024eqi,Lutfuoglu:2025bsf,Lutfuoglu:2026gis,Lutfuoglu:2026uzy}. At the same time, the emergence of an arbitrarily long-lived regime is not automatic, and there are examples in which increasing the field mass does not lead to vanishing damping \cite{Zinhailo:2024jzt}.

Massive fields also leave a clear imprint on the time-domain signal after the quasinormal ringing stage. Instead of the standard massless power-law tail, one typically finds oscillatory late-time behavior whose detailed form depends on both the field mass and the spacetime background \cite{Jing:2004zb,Koyama:2001qw,Moderski:2001tk,Rogatko:2007zz,Koyama:2001ee,Koyama:2000hj,Gibbons:2008gg,Gibbons:2008rs,Dubinsky:2024jqi}. A consistent analysis of massive perturbations therefore has to connect the shape of the effective potential, the behavior of the quasinormal spectrum, and the transition from ringdown to tails.

The same effective potential also governs the real-frequency scattering problem. In that context the grey-body factors (GBFs) and absorption cross sections (ACSs) measure how efficiently the massive scalar wave tunnels through the curvature barrier and is absorbed by the black hole. They therefore provide a natural complement to the quasinormal spectrum: while quasinormal modes probe the complex-frequency response of the geometry, GBFs and ACSs describe transmission and absorption for incident waves.

In this work we study quasinormal modes, GBFs, and ACSs of a massive scalar field in the RN-like geometry governed by the parameter $\gamma$. While perturations, scattering properties and spectrum of massless fields in Reissner-Nordstrom background is basically exhaustive \cite{Onozawa:1996ba,Onozawa:1995vu,Kokkotas:1988fm}, much less papers are dedicated to massive fields \cite{Konoplya:2002wt} while when taking into account the case of negative square charge almost all studies are concentrated on massless fields \cite{Malik:2024voy}. Our purpose is to determine how the charge-like deformation and the field mass modify the effective potential, the damping rate, the transmission probability, and the absorption cross section, and how they affect the onset of quasi-resonant behavior. To this end we combine Pad\'e-improved WKB calculations with time-domain integration for the quasinormal spectrum and the standard higher-order WKB approach for scattering, identify the parameter region in which the barrier-shaped potential necessary for the WKB treatment survives, and analyze how the resulting behavior depends on the multipole number and on $\gamma$.

The paper is organized as follows. In Sec.~\ref{sec:wavelike} we introduce the RN-like geometry, derive the exact scalar-field master equation, and discuss the effective potential together with the conditions under which it preserves a barrier shape. In Sec.~\ref{sec:WKB} we outline the semiclassical WKB treatment used for the quasinormal spectrum, while Sec.~\ref{sec:td} summarizes the time-domain integration method. Section~V presents the quasinormal-mode results, including the damping-rate fits, extrapolated quasi-resonant thresholds, and the comparison with the time-domain signal. In Sec.~\ref{sec:gbf} we study the grey-body factors and absorption cross sections, and in Sec.~VII we summarize the main conclusions.

\section{Background geometry and scalar-field master equation}\label{sec:wavelike}

We consider the static, spherically symmetric line element
\begin{equation}\label{metric}
  ds^2=-f(r)dt^2+\frac{dr^2}{f(r)}+r^2(d\theta^2+\sin^2\theta d\phi^2),
\end{equation}
with
\begin{equation}
f(r)=1-\frac{2 M}{r}-\frac{\gamma}{r^2}.
\end{equation}
This geometry is mathematically of the Reissner--Nordstr\"om type, but the interpretation of the $r^{-2}$ term need not be electromagnetic. In the braneworld construction of Dadhich, Maartens, Papadopoulos, and Rezania, the effective on-brane vacuum field equations admit the metric function $f(r)=1-2M/r-\gamma/r^2$, where $\gamma$ is a tidal charge induced by bulk gravitational effects rather than by a Maxwell field \cite{Dadhich:2000am}.  Thus the symbol $Q_{\rm eff}^2=-\gamma$ is only a formal Reissner--Nordstr\"om analogy: in the braneworld interpretation it should not be read as the square of a physical electric charge. In particular, $\gamma>0$ strengthens the gravitational field and places the event horizon outside the Schwarzschild value, while $\gamma<0$ gives a correction with the same sign as the usual Reissner--Nordstr\"om term; see also \cite{Seahra:2004fg}.  In what follows we consider massive scalar-field perturbations, quasinormal modes, grey-body factors, and absorption cross sections. Here $\mu$ is the mass of the scalar field, and we measure all dimensional quantities in units of the black-hole mass, that is, we set $M=1$.

We now derive the exact master equation for a massive scalar on the RN-like background (\ref{metric}), which is the equation used throughout the remainder of the paper.

The dynamics of the massive scalar field is governed by the Klein--Gordon equation,
\begin{equation}\label{KGmassive}
\frac{1}{\sqrt{-g}}\partial_\mu\left(\sqrt{-g}g^{\mu\nu}\partial_\nu\Phi\right)-\mu^2\Phi=0.
\end{equation}
Using the standard separation of variables \cite{Carter:1968ks,Konoplya:2018arm},
\begin{equation}
\Phi(t,r,\theta,\phi)=e^{-i\omega t}Y_{\ell m}(\theta,\phi)\phi_\ell(r),
\end{equation}
where $Y_{\ell m}(\theta,\phi)$ are the spherical harmonics, and the Klein--Gordon equation reduces to the radial equation
\begin{equation}
\begin{split}
\frac{f(r)}{r^2}\frac{d}{dr}\left(r^2f(r)\frac{d\phi_\ell}{dr}\right)\\
+\left(\omega^2-f(r)\left(\frac{\ell(\ell+1)}{r^2}+\mu^2\right)\right)\phi_\ell(r)=0.
\end{split}
\end{equation}
where $\ell=0,1,2,\ldots$ is the multipole number. Introducing the tortoise coordinate
\begin{equation}\label{tortoise}
dr_*\equiv\frac{dr}{f(r)}
\end{equation}
and redefining the radial function as
\begin{equation}
\phi_\ell(r)=\frac{\Psi(r)}{r},
\end{equation}
we reduce the perturbation equation to the Schr\"odinger-like master wave equation \cite{Kokkotas:1999bd,Berti:2009kk,Konoplya:2011qq}
\begin{equation}\label{wave-equation}
\dfrac{d^2\Psi}{dr_*^2}+(\omega^2-V(r))\Psi=0.
\end{equation}
The corresponding effective potential for the massive scalar field reads
\begin{equation}\label{potentialScalar}
\begin{split}
V(r)&=f(r)\left(\mu^2+\frac{\ell(\ell+1)}{r^2}+\frac{1}{r}\frac{df}{dr}\right)\\
&=f(r)\left(\mu^2+\frac{\ell(\ell+1)}{r^2}+\frac{2M}{r^3}+\frac{2\gamma}{r^4}\right).
\end{split}
\end{equation}
The effective potential vanishes at the event horizon and tends to $\mu^2$ at spatial infinity. Therefore, unlike the massless case, increasing the field mass lifts the asymptotic tail of the potential. For sufficiently small $\mu$ the curvature and angular-momentum terms still create a barrier peak, but as $\mu$ grows this peak is gradually washed out and eventually disappears.

\begin{figure}[t]
\centering
\includegraphics[width=\linewidth]{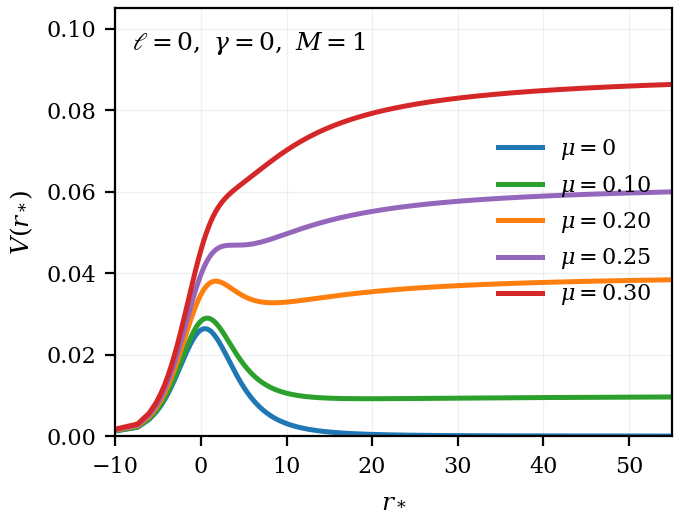}
\caption{Representative effective potentials as functions of the tortoise coordinate for the lowest multipole, $\ell=0$, in the Schwarzschild limit $\gamma=0$ with $M=1$. For small scalar-field masses the potential has a barrier peak. At the critical value $\mu=0.25$ the extremum degenerates, and for larger masses the potential becomes monotonic outside the horizon.}
\label{fig:potential-mu}
\end{figure}

This behavior is illustrated in Fig.~\ref{fig:potential-mu}. For the representative lowest multipole in the Schwarzschild limit, $\ell=0$ and $\gamma=0$, one has
\begin{equation}
V(r)=\left(1-\frac{2}{r}\right)\left(\mu^2+\frac{2}{r^3}\right),
\end{equation}
and the extremum condition $dV/dr=0$ reduces to $2\mu^2r^3-6r+16=0$. The critical configuration is reached when the maximum and minimum merge into a single inflection point, which gives $r=4$ and $\mu M=1/4$; in our units this means $\mu=0.25$. Thus, for $\mu<0.25$ the effective potential has the barrier shape required by the WKB approximation, whereas for $\mu>0.25$ it becomes monotonic outside the horizon. Although the precise critical value depends on $\gamma$ and $\ell$, the same qualitative behavior persists throughout the parameter space.

\begin{figure}[t]
\centering
\includegraphics[width=\linewidth]{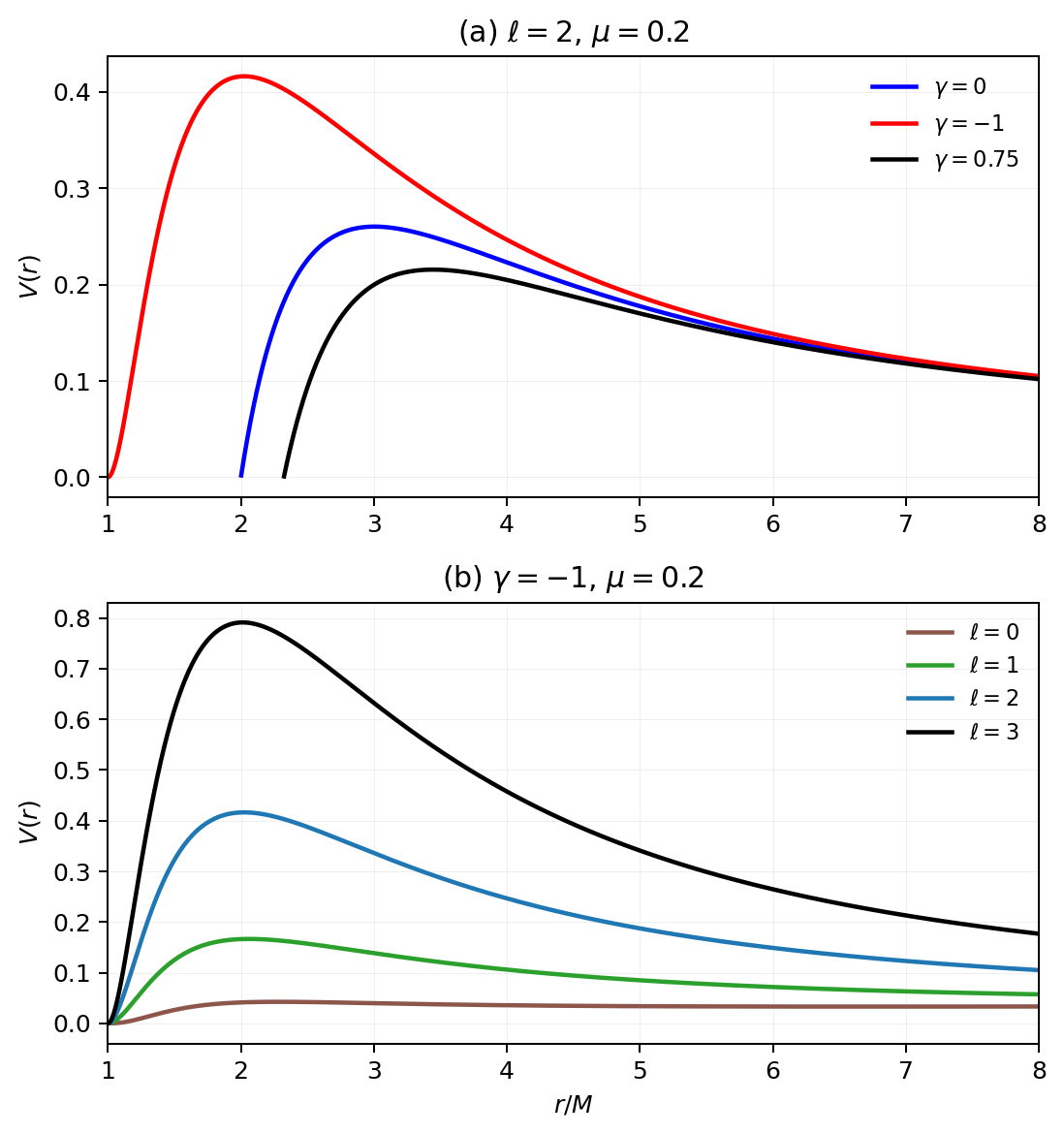}
\caption{Representative effective potentials as functions of the radial coordinate. Panel (a) shows the dependence on the tidal-charge parameter for $\ell=2$ and $\mu=0.2$, while panel (b) shows the dependence on the multipole number for $\gamma=-1$ and $\mu=0.2$.}
\label{fig:potential-params}
\end{figure}

To connect the scattering results below with the barrier geometry, Figs.~\ref{fig:potential-mu} and \ref{fig:potential-params} show three key trends. First, increasing $\mu$ lifts the asymptotic tail and, for the representative cases relevant to Sec.~\ref{sec:gbf}, also raises the barrier peak. Second, increasing $\ell$ rapidly enhances the barrier through the centrifugal term. Third, increasing $\gamma$ lowers the peak, whereas more negative $\gamma$ raises it. These trends already indicate that a higher barrier should suppress transmission and therefore reduce the grey-body factors.

This behavior is important for the analysis below. The WKB treatment of quasinormal modes and grey-body factors assumes a single barrier-shaped potential, so parameter values for which the peak disappears must be excluded from the WKB tables and treated instead with methods such as direct time-domain evolution.

\section{Semiclassical WKB approach}\label{sec:WKB}
For each multipole number the radial equation (\ref{wave-equation}) can be interpreted as a one-dimensional scattering problem written in terms of the tortoise coordinate. Quasinormal modes are obtained by demanding that no signal emerges from the event horizon and that only the outgoing component is present at spatial infinity. For the massive scalar field this gives
\begin{equation}\label{boundaryconditions}
\Psi\sim e^{-\imo\omega r_*}, \quad r_*\to-\infty,
\qquad
\Psi\sim e^{+\imo\chi r_*}, \quad r_*\to+\infty,
\end{equation}
where
\begin{equation}
\chi=\sqrt{\omega^2-\mu^2},
\end{equation}
and the branch of $\chi$ is fixed by the requirement of an outgoing wave at infinity.

The WKB construction becomes especially useful when the effective potential is a single smooth barrier. Instead of solving the wave equation globally, one expands the potential about its maximum at $r=r_{\rm max}$, finds the local solution in this region, and matches it to the WKB solutions on the two sides of the barrier. In this way the frequency is expressed entirely through the derivatives of the potential at the peak,
\begin{equation}
V_j\equiv\left.\frac{d^jV}{dr_*^j}\right|_{r_*=r_*^{\rm max}},
\end{equation}
with $V_1=0$ and $V_2<0$. The resulting asymptotic expansion can be written as \cite{Konoplya:2019hlu}
\begin{equation}\label{WKBformula-spherical}
\begin{split}
\omega^2={}&V_0+A_2(\K^2)+A_4(\K^2)+A_6(\K^2)+\ldots\\
&-\imo\K\sqrt{-2V_2}\\
&\times\left(1+A_3(\K^2)+A_5(\K^2)+A_7(\K^2)+\ldots\right).
\end{split}
\end{equation}
where
\begin{equation}
\K=n+\frac{1}{2}, \qquad n=0,1,2,\ldots.
\end{equation}

The first term represents the eikonal limit, whereas the functions $A_j$ supply successive finite-wavelength corrections. Each $A_j$ is an algebraic expression depending on $\K$ and on the derivatives $V_j$ at the maximum, with higher-order terms requiring progressively higher derivatives of the potential. Since these expressions rapidly become cumbersome, we do not reproduce them here; representative derivations are given in \cite{Iyer:1986np,Konoplya:2003ii,Matyjasek:2017psv}. In the calculations below we use Pad\'e-improved WKB series, because Pad\'e resummation usually makes the truncated expansion considerably more stable for the low-lying modes. For the grey-body factors discussed later we keep the same WKB framework, but there we use the standard sixth-order WKB approximation for the scattering problem rather than a Pad\'e-resummed scheme. Throughout this work we apply it only in the parameter region where the potential preserves a peak barrier shape and the approximation is expected to remain trustworthy \cite{Guo:2020caw,Bolokhov:2025aqy,Konoplya:2009hv,Malik:2025erb,Kokkotas:2010zd,Konoplya:2005sy,Eniceicu:2019npi,Bolokhov:2025egl,Malik:2026lfj,Gonzalez:2022ote,Pathrikar:2025gzu,Fernando:2016ftj,Bolokhov:2025lnt,Momennia:2018hsm,Konoplya:2019hlu}.

\section{Time-domain integration}\label{sec:td}

An independent way to analyze the response of the black hole to massive scalar perturbations is to evolve the master equation directly in the time domain. Unlike the WKB method, which uses local information near the maximum of the effective potential, the time-domain approach follows the full wave packet and therefore reveals the prompt response, the quasinormal ringing stage, and the late-time tail within a single computation.

For this purpose it is convenient to introduce the light-cone coordinates
\begin{equation}
u=t-r_*, \qquad v=t+r_*,
\end{equation}
in terms of which the wave equation (\ref{wave-equation}) takes the form
\begin{equation}
4\frac{\partial^2\Psi}{\partial u\partial v}+V(u,v)\Psi=0,
\end{equation}
where $V(u,v)=V(r(u,v))$. We then discretize the $(u,v)$ plane by a uniform null grid with step $\Delta$ and consider the four points of a null rectangle,
\begin{equation}
\begin{split}
S=(u,v), \qquad E=(u,v+\Delta),\\
W=(u+\Delta,v), \qquad N=(u+\Delta,v+\Delta).
\end{split}
\end{equation}
Expanding the field and the potential inside this cell, Gundlach, Price, and Pullin obtained the characteristic integration scheme \cite{Gundlach:1993tp}
\begin{eqnarray}
\Psi\left(N\right)&=&\Psi\left(W\right)+\Psi\left(E\right)-\Psi\left(S\right)\nonumber\\
&&-\frac{\Delta^2}{8}V\left(S\right)\left[\Psi\left(W\right)+\Psi\left(E\right)\right]+{\cal O}\left(\Delta^4\right),\label{Discretization}
\end{eqnarray}
which is second-order accurate in the grid spacing and is straightforward to implement numerically.

The evolution starts from characteristic initial data prescribed on two null segments, for example a Gaussian pulse on one of them and a trivial profile on the other. Repeated use of the update formula (\ref{Discretization}) determines the scalar field at all subsequent grid points. The waveform recorded at a fixed radial position typically consists of three stages: an early-time burst sensitive to the initial data, an intermediate interval dominated by damped oscillations of the form $e^{-i\omega t}$, and a late-time tail. The complex quasinormal frequency can therefore be extracted from the ringing part of the profile, providing an important check of the WKB results obtained from the frequency-domain analysis.

This characteristic integration scheme has become a standard tool in black-hole perturbation theory and has been applied successfully in many related problems\cite{Skvortsova:2024atk,Konoplya:2023aph,Arbelaez:2026eaz,Malik:2024nhy,Bolokhov:2024ixe,Varghese:2011ku,Konoplya:2014lha,Bolokhov:2023dxq,Konoplya:2024kih,Skvortsova:2024wly,Abdalla:2012si,Konoplya:2005et,Aneesh:2018hlp,Malik:2023bxc,Konoplya:2024hfg,Bolokhov:2023bwm,Arbelaez:2025gwj,Skvortsova:2023zmj,Ishihara:2008re,Qian:2022kaq,Dubinsky:2025fwv,Varghese:2011ku,Konoplya:2018yrp}. In the present context it is especially useful because it does not rely on the existence of a sharp WKB barrier approximation alone, and it also allows one to follow the transition from quasinormal ringing to the asymptotic tail regime for massive scalar fields.

\section{Quasinormal modes}

Before listing the frequency tables, it is useful to isolate the dependence of the damping rate,
\begin{equation}
\omega_I\equiv-\im\omega,
\end{equation}
on the scalar-field mass.  For each selected $(\gamma,\ell,n)$ family we fitted the last six available WKB16 points by a quadratic polynomial,
\begin{equation}
\omega_I(\mu)\simeq a+b\mu+c\mu^2,
\end{equation}
and defined the critical mass $\mu_c$ as the positive root of $\omega_I(\mu_c)=0$. Since the WKB data terminate when the potential barrier ceases to be reliable, the extracted $\mu_c$ should be understood as an extrapolated quasi-resonant threshold rather than as a directly computed WKB point.

\begin{figure*}[t]
\centering
\includegraphics[width=0.9\textwidth]{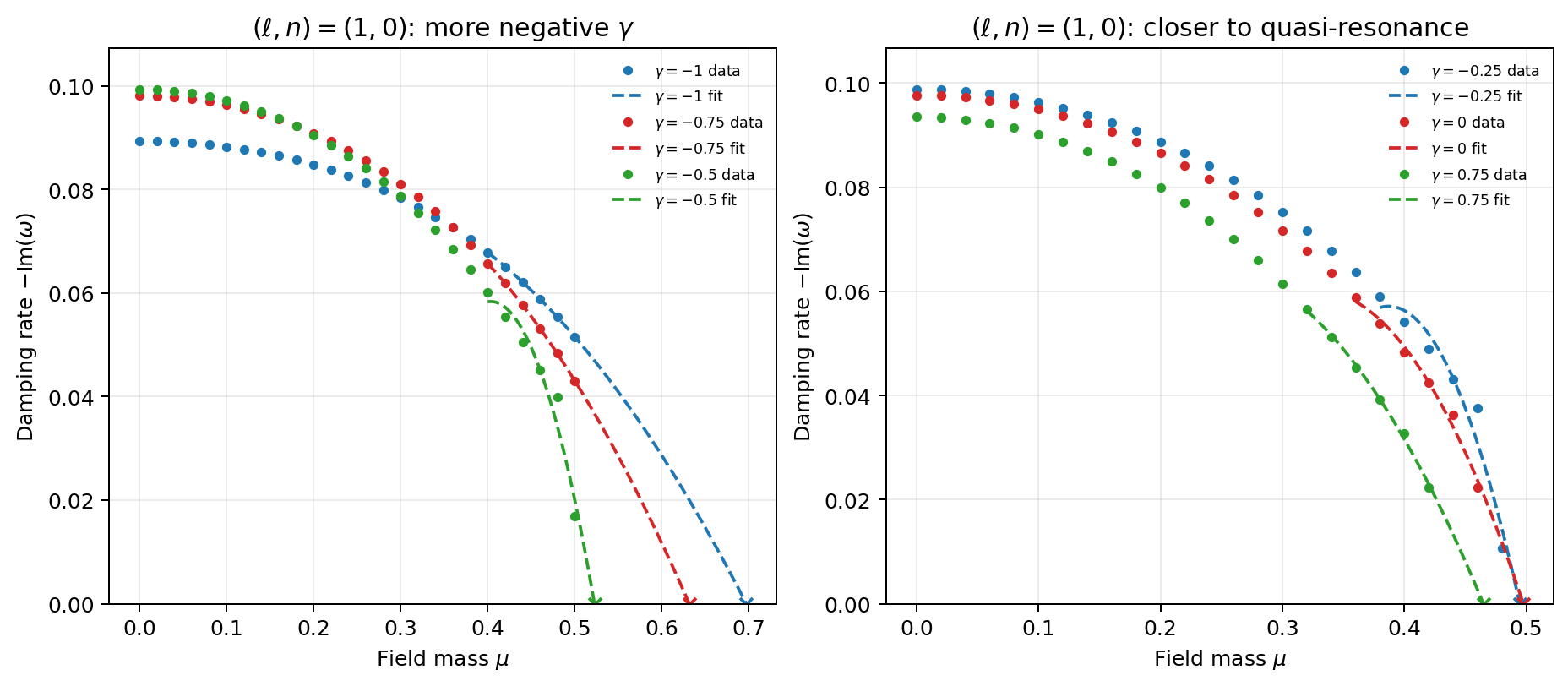}
\caption{Damping rate $\omega_I=-\im\omega$ as a function of the scalar-field mass for selected quasinormal-mode families. Circular markers show the WKB16 data extracted from the full mass grids contained in the source tables, dashed curves are quadratic fits to the last six available points, and crosses mark the extrapolated critical masses $\mu_c$ at which the damping rate vanishes. Left: fundamental modes with $(\ell,n)=(1,0)$ for $\gamma=-1$, $-0.75$, and $-0.5$. Right: the same mode family for $\gamma=-0.25$, $0$, and $0.75$, chosen because these cases approach the quasi-resonant regime more closely before the WKB barrier disappears.}
\label{fig:damping-mu}
\end{figure*}

Figure~\ref{fig:damping-mu} focuses on the fundamental family $(\ell,n)=(1,0)$, because these modes provide the clearest WKB illustration of the approach to quasi-resonance. The left panel displays more negative values of the tidal-charge parameter, $\gamma=-1$, $-0.75$, and $-0.5$, while the right panel shows $\gamma=-0.25$, $0$, and $0.75$, for which the last WKB points lie much closer to the horizontal axis. In this way the second panel makes the tendency $\omega_I\to0$ with increasing $\mu$ visually explicit, while still remaining within the parameter region where a barrier peak is present.

\begin{table}[t]
\centering
\caption{Extrapolated critical masses $\mu_c$ obtained from quadratic fits to the damping-rate tails shown in Fig.~\ref{fig:damping-mu}. The fit is performed on the last six available WKB16 points of each selected fundamental $(\ell,n)=(1,0)$ family.}
\label{tab:mu-critical}
\begin{tabular}{c c c c}
\hline
$\ell$ & $n$ & $\gamma$ & $\mu_c$ \\
\hline
1 & 0 & $-1.00$ & $0.697$ \\
1 & 0 & $-0.75$ & $0.632$ \\
1 & 0 & $-0.50$ & $0.523$ \\
1 & 0 & $-0.25$ & $0.495$ \\
1 & 0 & $0.00$ & $0.498$ \\
1 & 0 & $0.75$ & $0.465$ \\
\hline
\end{tabular}
\end{table}

These extrapolations show that the quasi-resonant threshold is distinct from the geometric condition for the disappearance of the potential barrier discussed in Sec.~\ref{sec:wavelike}. For example, in the Schwarzschild $\ell=0$ case the barrier itself disappears already at $\mu=0.25$, while the $\ell=1$ fundamental modes shown here preserve a barrier up to substantially larger masses and allow one to follow the damping rate deep into the quasi-resonant regime. For the selected families the extrapolated threshold lies in the range $\mu_c\simeq0.465$--$0.697$. The right panel of Fig.~\ref{fig:damping-mu} is particularly useful in this respect, because it isolates the parameter choices for which the available WKB data come visibly closest to zero damping before the barrier-shaped potential is lost.

\begin{figure}[!t]
\centering
\includegraphics[width=\linewidth]{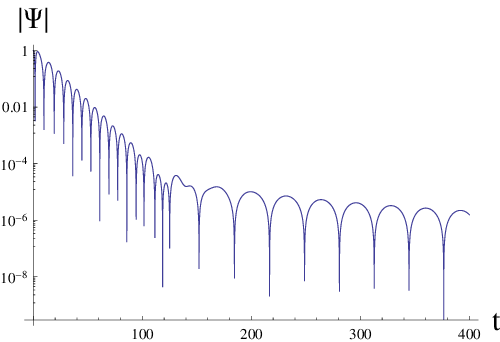}
\caption{Time-domain profile $|\Psi|$ for the massive scalar perturbation of the RN-like black hole with $M=1$, $\ell=1$, $\mu=0.1$, and $\gamma=-1$. The intermediate ringing stage is fitted by the fundamental mode extracted with the Prony method, $\omega=0.380626-0.0882639 i$, in excellent agreement with the WKB16 value from Table~\ref{tab:qnm-wkb}, panel (b), at the level of $0.0038\%$.}
\label{fig:td-profile}
\end{figure}

As an explicit time-domain check for the fundamental family, Fig.~\ref{fig:td-profile} shows the waveform for $M=1$, $\ell=1$, $\mu=0.1$, and $\gamma=-1$. A Prony fit to the ringing stage gives $\omega_{\rm Prony}=0.380626-0.0882639 i$. The corresponding WKB16 value from Table~\ref{tab:qnm-wkb}, panel (b), is $\omega_{\rm WKB}=0.380631-0.088278 i$, so the relative difference $100|\omega_{\rm Prony}-\omega_{\rm WKB}|/|\omega_{\rm WKB}|$ is only $0.0038\%$.

\begin{table*}[!t]
\centering
\begin{minipage}[t]{0.49\textwidth}
\centering
\textbf{(a) $\ell=0$, $n=0$}\\[2pt]
{\scriptsize
\renewcommand{\arraystretch}{0.95}
\setlength{\tabcolsep}{3pt}
\resizebox{\linewidth}{!}{%
\begin{tabular}{c c c c c}
\hline
$\gamma$ & $\mu$ & WKB16 ($\tilde{m}=8$) & WKB14 ($\tilde{m}=7$) & difference  \\
\hline
$-1.$ & $0$ & $0.133838-0.096059 i$ & $0.133350-0.095632 i$ & $0.393\%$\\
$-1.$ & $0.08$ & $0.134710-0.092804 i$ & $0.135715-0.092043 i$ & $0.771\%$\\
$-1.$ & $0.16$ & $0.141551-0.081872 i$ & $0.142766-0.082538 i$ & $0.847\%$\\
$-1.$ & $0.22$ & $0.161286-0.071120 i$ & $0.146652-0.068277 i$ & $8.46\%$\\
$-1.$ & $0.24$ & $0.144865-0.063945 i$ & $0.142737-0.064797 i$ & $1.45\%$\\
$-0.75$ & $0$ & $0.130231-0.103559 i$ & $0.130234-0.103522 i$ & $0.0228\%$\\
$-0.75$ & $0.08$ & $0.131843-0.099664 i$ & $0.131761-0.099487 i$ & $0.118\%$\\
$-0.75$ & $0.16$ & $0.136453-0.088410 i$ & $0.136819-0.088655 i$ & $0.271\%$\\
$-0.75$ & $0.2$ & $0.140495-0.079921 i$ & $0.137536-0.081385 i$ & $2.04\%$\\
$-0.75$ & $0.24$ & $0.135187-0.070756 i$ & $0.133759-0.076344 i$ & $3.78\%$\\
$-0.5$ & $0$ & $0.122361-0.105882 i$ & $0.122180-0.105516 i$ & $0.252\%$\\
$-0.5$ & $0.08$ & $0.123979-0.101432 i$ & $0.123513-0.100790 i$ & $0.495\%$\\
$-0.5$ & $0.16$ & $0.127772-0.089113 i$ & $0.127883-0.088359 i$ & $0.489\%$\\
$-0.5$ & $0.18$ & $0.128660-0.082840 i$ & $0.127655-0.085745 i$ & $2.01\%$\\
$-0.5$ & $0.2$ & $0.127100-0.082073 i$ & $0.123940-0.084789 i$ & $2.75\%$\\
$-0.5$ & $0.24$ & $0.126047-0.085031 i$ & $0.132765-0.089586 i$ & $5.34\%$\\
$-0.25$ & $0$ & $0.115812-0.105814 i$ & $0.115692-0.105476 i$ & $0.229\%$\\
$-0.25$ & $0.08$ & $0.117348-0.101043 i$ & $0.116826-0.100302 i$ & $0.585\%$\\
$-0.25$ & $0.1$ & $0.118220-0.098240 i$ & $0.119955-0.095492 i$ & $2.11\%$\\
$-0.25$ & $0.16$ & $0.120915-0.087822 i$ & $0.120986-0.087040 i$ & $0.525\%$\\
$-0.25$ & $0.18$ & $0.121326-0.081334 i$ & $0.120271-0.085879 i$ & $3.19\%$\\
$-0.25$ & $0.2$ & $0.118995-0.091657 i$ & $0.124334-0.074322 i$ & $12.1\%$\\
$-0.25$ & $0.24$ & $0.139608-0.090268 i$ & $0.148647-0.085915 i$ & $6.03\%$\\
$0$ & $0$ & $0.110473-0.104954 i$ & $0.110394-0.104612 i$ & $0.230\%$\\
$0$ & $0.08$ & $0.111937-0.099955 i$ & $0.111366-0.099000 i$ & $0.741\%$\\
$0$ & $0.1$ & $0.112880-0.096990 i$ & $0.115824-0.095846 i$ & $2.12\%$\\
$0$ & $0.16$ & $0.115319-0.086315 i$ & $0.115262-0.085426 i$ & $0.618\%$\\
$0$ & $0.2$ & $0.125466-0.067499 i$ & $0.114439-0.074719 i$ & $9.25\%$\\
$0$ & $0.22$ & $0.110113-0.071318 i$ & $0.108140-0.074331 i$ & $2.74\%$\\
$0$ & $0.24$ & $0.160768-0.065715 i$ & $0.164864-0.057826 i$ & $5.12\%$\\
$0.25$ & $0$ & $0.106004-0.103776 i$ & $0.105959-0.103437 i$ & $0.231\%$\\
$0.25$ & $0.08$ & $0.107403-0.098592 i$ & $0.106765-0.097410 i$ & $0.921\%$\\
$0.25$ & $0.1$ & $0.108429-0.095472 i$ & $0.110971-0.095307 i$ & $1.76\%$\\
$0.25$ & $0.16$ & $0.110732-0.085212 i$ & $0.110357-0.083788 i$ & $1.05\%$\\
$0.25$ & $0.2$ & $0.110204-0.071173 i$ & $0.107460-0.073198 i$ & $2.60\%$\\
$0.25$ & $0.22$ & $0.101138-0.074931 i$ & $0.103057-0.082707 i$ & $6.36\%$\\
$0.25$ & $0.24$ & $0.180096-0.021352 i$ & $0.183052-0.017213 i$ & $2.80\%$\\
$0.5$ & $0$ & $0.102178-0.102475 i$ & $0.102161-0.102138 i$ & $0.233\%$\\
$0.5$ & $0.08$ & $0.103521-0.097148 i$ & $0.102797-0.095693 i$ & $1.15\%$\\
$0.5$ & $0.16$ & $0.108370-0.085271 i$ & $0.105988-0.082292 i$ & $2.77\%$\\
$0.5$ & $0.18$ & $0.103632-0.082195 i$ & $0.104160-0.086172 i$ & $3.03\%$\\
$0.5$ & $0.22$ & $0.103053-0.087104 i$ & $0.113367-0.090786 i$ & $8.12\%$\\
$0.75$ & $0$ & $0.098844-0.101138 i$ & $0.098851-0.100799 i$ & $0.240\%$\\
$0.75$ & $0.08$ & $0.100146-0.095703 i$ & $0.099321-0.093896 i$ & $1.43\%$\\
$0.75$ & $0.16$ & $0.109738-0.079478 i$ & $0.102005-0.080960 i$ & $5.81\%$\\
$0.75$ & $0.18$ & $0.099392-0.081182 i$ & $0.119213-0.076361 i$ & $15.9\%$\\
$0.75$ & $0.22$ & $0.122962-0.087907 i$ & $0.132569-0.082519 i$ & $7.29\%$\\
\hline
\end{tabular}%
}}
\end{minipage}\hfill
\begin{minipage}[t]{0.49\textwidth}
\centering
\textbf{(b) $\ell=1$, $n=0$}\\[2pt]
{\scriptsize
\renewcommand{\arraystretch}{0.95}
\setlength{\tabcolsep}{3pt}
\resizebox{\linewidth}{!}{%
\begin{tabular}{c c c c c}
\hline
$\gamma$  & $\mu$ & WKB16 ($\tilde{m}=8$) & WKB14 ($\tilde{m}=7$) & difference  \\
\hline
$-1.$ & $0$ & $0.377642-0.089384 i$ & $0.377642-0.089384 i$ & $0\%$\\
$-1.$ & $0.1$ & $0.380631-0.088278 i$ & $0.380631-0.088278 i$ & $0\%$\\
$-1.$ & $0.2$ & $0.389653-0.084790 i$ & $0.389653-0.084790 i$ & $0\%$\\
$-1.$ & $0.3$ & $0.404844-0.078356 i$ & $0.404844-0.078356 i$ & $0\%$\\
$-1.$ & $0.4$ & $0.426243-0.067843 i$ & $0.426243-0.067844 i$ & $0.00012\%$\\
$-1.$ & $0.5$ & $0.453183-0.051546 i$ & $0.453175-0.051570 i$ & $0.00549\%$\\
$-0.75$ & $0$ & $0.345964-0.098132 i$ & $0.345964-0.098132 i$ & $0\%$\\
$-0.75$ & $0.1$ & $0.349428-0.096367 i$ & $0.349428-0.096367 i$ & $0\%$\\
$-0.75$ & $0.2$ & $0.359883-0.090896 i$ & $0.359883-0.090896 i$ & $0.\times 10^{\text{-4}}\%$\\
$-0.75$ & $0.3$ & $0.377470-0.081106 i$ & $0.377470-0.081106 i$ & $0\%$\\
$-0.75$ & $0.4$ & $0.402145-0.065750 i$ & $0.402145-0.065752 i$ & $0.00064\%$\\
$-0.75$ & $0.5$ & $0.432594-0.043001 i$ & $0.432391-0.043374 i$ & $0.0977\%$\\
$-0.5$ & $0$ & $0.323534-0.099352 i$ & $0.323534-0.099352 i$ & $0\%$\\
$-0.5$ & $0.1$ & $0.327397-0.097199 i$ & $0.327397-0.097199 i$ & $0\%$\\
$-0.5$ & $0.2$ & $0.339057-0.090538 i$ & $0.339057-0.090538 i$ & $0\%$\\
$-0.5$ & $0.3$ & $0.358672-0.078663 i$ & $0.358672-0.078663 i$ & $0.\times 10^{\text{-4}}\%$\\
$-0.5$ & $0.4$ & $0.386117-0.060138 i$ & $0.386110-0.060127 i$ & $0.00340\%$\\
$-0.5$ & $0.5$ & $0.443564-0.016966 i$ & $0.444999-0.014320 i$ & $0.678\%$\\
$-0.25$ & $0$ & $0.306562-0.098798 i$ & $0.306562-0.098798 i$ & $0\%$\\
$-0.25$ & $0.1$ & $0.310755-0.096347 i$ & $0.310755-0.096347 i$ & $0\%$\\
$-0.25$ & $0.2$ & $0.323419-0.088764 i$ & $0.323419-0.088764 i$ & $0\%$\\
$-0.25$ & $0.3$ & $0.344739-0.075244 i$ & $0.344739-0.075244 i$ & $0.\times 10^{\text{-4}}\%$\\
$-0.25$ & $0.4$ & $0.374520-0.054124 i$ & $0.374501-0.054149 i$ & $0.00851\%$\\
$-0.25$ & $0.48$ & $0.430750-0.010696 i$ & $0.432176-0.008733 i$ & $0.563\%$\\
$0$ & $0$ & $0.292936-0.097660 i$ & $0.292936-0.097660 i$ & $0\%$\\
$0$ & $0.1$ & $0.297416-0.094957 i$ & $0.297416-0.094957 i$ & $0\%$\\
$0$ & $0.2$ & $0.310957-0.086593 i$ & $0.310957-0.086593 i$ & $0\%$\\
$0$ & $0.3$ & $0.333777-0.071658 i$ & $0.333778-0.071658 i$ & $0.\times 10^{\text{-4}}\%$\\
$0$ & $0.4$ & $0.365599-0.048285 i$ & $0.365587-0.048280 i$ & $0.00343\%$\\
$0$ & $0.46$ & $0.408065-0.022383 i$ & $0.409747-0.019507 i$ & $0.815\%$\\
$0.25$ & $0$ & $0.281573-0.096314 i$ & $0.281573-0.096314 i$ & $0\%$\\
$0.25$ & $0.1$ & $0.286310-0.093389 i$ & $0.286310-0.093389 i$ & $0\%$\\
$0.25$ & $0.2$ & $0.300640-0.084334 i$ & $0.300640-0.084334 i$ & $0\%$\\
$0.25$ & $0.3$ & $0.324818-0.068130 i$ & $0.324818-0.068130 i$ & $0\%$\\
$0.25$ & $0.4$ & $0.358437-0.042706 i$ & $0.358401-0.042734 i$ & $0.0128\%$\\
$0.25$ & $0.44$ & $0.380462-0.033908 i$ & $0.383367-0.032918 i$ & $0.803\%$\\
$0.5$ & $0$ & $0.271846-0.094906 i$ & $0.271846-0.094906 i$ & $0\%$\\
$0.5$ & $0.1$ & $0.276817-0.091782 i$ & $0.276817-0.091782 i$ & $0\%$\\
$0.5$ & $0.2$ & $0.291872-0.082099 i$ & $0.291872-0.082099 i$ & $0\%$\\
$0.5$ & $0.3$ & $0.317302-0.064728 i$ & $0.317302-0.064728 i$ & $0.\times 10^{\text{-4}}\%$\\
$0.5$ & $0.4$ & $0.352485-0.037507 i$ & $0.352405-0.037625 i$ & $0.0403\%$\\
$0.5$ & $0.42$ & $0.360158-0.032031 i$ & $0.360679-0.032985 i$ & $0.301\%$\\
$0.75$ & $0$ & $0.263357-0.093503 i$ & $0.263357-0.093503 i$ & $0\%$\\
$0.75$ & $0.1$ & $0.268546-0.090195 i$ & $0.268546-0.090195 i$ & $0\%$\\
$0.75$ & $0.2$ & $0.284273-0.079933 i$ & $0.284273-0.079933 i$ & $0\%$\\
$0.75$ & $0.3$ & $0.310873-0.061472 i$ & $0.310873-0.061472 i$ & $0.\times 10^{\text{-4}}\%$\\
$0.75$ & $0.4$ & $0.347456-0.032681 i$ & $0.347415-0.032736 i$ & $0.0197\%$\\
$0.75$ & $0.42$ & $0.373817-0.022330 i$ & $0.375763-0.019671 i$ & $0.880\%$\\
\hline
\end{tabular}%
}}
\end{minipage}
\caption{Representative fundamental scalar quasinormal frequencies of the Reissner--Nordstr\"om black hole with $M=1$, computed with the Pad\'e-improved 16th- and 14th-order WKB formulas. Panel (a) shows the $\ell=0$, $n=0$ mode, while panel (b) shows the $\ell=1$, $n=0$ mode. Rows with only weak frequency variation are omitted from the printed tables and remain commented in the source, and cases with no barrier peak are removed. The last column gives the relative percent difference between the two WKB estimates.}
\label{tab:qnm-wkb}
\end{table*}

\begin{table*}[!t]
\centering
\begin{minipage}[t]{0.49\textwidth}
\centering
\textbf{(a) $\ell=2$, $n=0$}\\[2pt]
{\scriptsize
\renewcommand{\arraystretch}{0.95}
\setlength{\tabcolsep}{3pt}
\resizebox{\linewidth}{!}{%
\begin{tabular}{c c c c c}
\hline
$\gamma$  & $\mu$ & WKB16 ($\tilde{m}=8$) & WKB14 ($\tilde{m}=7$) & difference  \\
\hline
$-1.$ & $0$ & $0.626573-0.088748 i$ & $0.626573-0.088748 i$ & $0\%$\\
$-1.$ & $0.08$ & $0.627804-0.088483 i$ & $0.627804-0.088483 i$ & $0\%$\\
$-1.$ & $0.16$ & $0.631508-0.087677 i$ & $0.631508-0.087677 i$ & $0\%$\\
$-1.$ & $0.24$ & $0.637720-0.086298 i$ & $0.637720-0.086298 i$ & $0\%$\\
$-0.75$ & $0$ & $0.570878-0.097507 i$ & $0.570878-0.097507 i$ & $0\%$\\
$-0.75$ & $0.08$ & $0.572390-0.097062 i$ & $0.572390-0.097062 i$ & $0\%$\\
$-0.75$ & $0.16$ & $0.576938-0.095718 i$ & $0.576938-0.095718 i$ & $0\%$\\
$-0.75$ & $0.24$ & $0.584560-0.093442 i$ & $0.584560-0.093442 i$ & $0\%$\\
$-0.5$ & $0$ & $0.533818-0.098572 i$ & $0.533818-0.098572 i$ & $0\%$\\
$-0.5$ & $0.08$ & $0.535533-0.098025 i$ & $0.535533-0.098025 i$ & $0\%$\\
$-0.5$ & $0.16$ & $0.540690-0.096372 i$ & $0.540690-0.096372 i$ & $0\%$\\
$-0.5$ & $0.24$ & $0.549333-0.093575 i$ & $0.549333-0.093575 i$ & $0\%$\\
$-0.25$ & $0$ & $0.505968-0.097942 i$ & $0.505968-0.097942 i$ & $0\%$\\
$-0.25$ & $0.08$ & $0.507847-0.097316 i$ & $0.507847-0.097316 i$ & $0\%$\\
$-0.25$ & $0.16$ & $0.513501-0.095423 i$ & $0.513501-0.095423 i$ & $0\%$\\
$-0.25$ & $0.24$ & $0.522976-0.092227 i$ & $0.522976-0.092227 i$ & $0\%$\\
$0$ & $0$ & $0.483644-0.096759 i$ & $0.483644-0.096759 i$ & $0\%$\\
$0$ & $0.08$ & $0.485665-0.096066 i$ & $0.485665-0.096066 i$ & $0\%$\\
$0$ & $0.16$ & $0.491747-0.093972 i$ & $0.491747-0.093972 i$ & $0\%$\\
$0$ & $0.24$ & $0.501944-0.090438 i$ & $0.501944-0.090438 i$ & $0\%$\\
$0.25$ & $0$ & $0.465032-0.095383 i$ & $0.465032-0.095383 i$ & $0\%$\\
$0.25$ & $0.08$ & $0.467181-0.094631 i$ & $0.467181-0.094631 i$ & $0\%$\\
$0.25$ & $0.16$ & $0.473646-0.092361 i$ & $0.473646-0.092361 i$ & $0\%$\\
$0.25$ & $0.24$ & $0.484487-0.088529 i$ & $0.484487-0.088529 i$ & $0\%$\\
$0.5$ & $0$ & $0.449097-0.093957 i$ & $0.449097-0.093957 i$ & $0\%$\\
$0.5$ & $0.08$ & $0.451361-0.093151 i$ & $0.451361-0.093151 i$ & $0\%$\\
$0.5$ & $0.16$ & $0.458175-0.090721 i$ & $0.458175-0.090721 i$ & $0\%$\\
$0.5$ & $0.24$ & $0.469606-0.086620 i$ & $0.469606-0.086620 i$ & $0\%$\\
$0.75$ & $0$ & $0.435185-0.092540 i$ & $0.435185-0.092540 i$ & $0\%$\\
$0.75$ & $0.08$ & $0.437556-0.091686 i$ & $0.437556-0.091686 i$ & $0\%$\\
$0.75$ & $0.16$ & $0.444694-0.089109 i$ & $0.444694-0.089109 i$ & $0\%$\\
$0.75$ & $0.24$ & $0.456670-0.084759 i$ & $0.456670-0.084759 i$ & $0\%$\\
\hline
\end{tabular}%
}}
\end{minipage}\hfill
\begin{minipage}[t]{0.49\textwidth}
\centering
\textbf{(b) $\ell=2$, $n=1$}\\[2pt]
{\scriptsize
\renewcommand{\arraystretch}{0.95}
\setlength{\tabcolsep}{3pt}
\resizebox{\linewidth}{!}{%
\begin{tabular}{c c c c c}
\hline
$\gamma$  & $\mu$ & WKB16 ($\tilde{m}=8$) & WKB14 ($\tilde{m}=7$) & difference  \\
\hline
$-1.$ & $0$ & $0.608171-0.269093 i$ & $0.608171-0.269093 i$ & $0\%$\\
$-1.$ & $0.1$ & $0.609662-0.267997 i$ & $0.609662-0.267997 i$ & $0\%$\\
$-1.$ & $0.2$ & $0.614112-0.264673 i$ & $0.614112-0.264673 i$ & $0\%$\\
$-1.$ & $0.3$ & $0.621451-0.259008 i$ & $0.621451-0.259008 i$ & $0\%$\\
$-1.$ & $0.4$ & $0.631538-0.250810 i$ & $0.631537-0.250810 i$ & $0\%$\\
$-1.$ & $0.5$ & $0.644119-0.239819 i$ & $0.644119-0.239819 i$ & $0.00002\%$\\
$-0.75$ & $0$ & $0.555984-0.296058 i$ & $0.555984-0.296058 i$ & $0\%$\\
$-0.75$ & $0.1$ & $0.557431-0.294458 i$ & $0.557431-0.294458 i$ & $0\%$\\
$-0.75$ & $0.2$ & $0.561747-0.289616 i$ & $0.561747-0.289616 i$ & $0\%$\\
$-0.75$ & $0.3$ & $0.568851-0.281409 i$ & $0.568851-0.281409 i$ & $0\%$\\
$-0.75$ & $0.4$ & $0.578573-0.269630 i$ & $0.578573-0.269630 i$ & $0.00001\%$\\
$-0.75$ & $0.5$ & $0.590594-0.254016 i$ & $0.590594-0.254016 i$ & $0.00006\%$\\
$-0.5$ & $0$ & $0.516782-0.300190 i$ & $0.516782-0.300190 i$ & $0\%$\\
$-0.5$ & $0.1$ & $0.518287-0.298281 i$ & $0.518287-0.298281 i$ & $0\%$\\
$-0.5$ & $0.2$ & $0.522773-0.292506 i$ & $0.522773-0.292507 i$ & $0\%$\\
$-0.5$ & $0.3$ & $0.530137-0.282721 i$ & $0.530137-0.282721 i$ & $0\%$\\
$-0.5$ & $0.4$ & $0.540163-0.268687 i$ & $0.540163-0.268687 i$ & $0\%$\\
$-0.5$ & $0.5$ & $0.552442-0.250118 i$ & $0.552442-0.250118 i$ & $0.00008\%$\\
$-0.25$ & $0$ & $0.487307-0.298817 i$ & $0.487307-0.298817 i$ & $0\%$\\
$-0.25$ & $0.1$ & $0.488871-0.296674 i$ & $0.488871-0.296674 i$ & $0\%$\\
$-0.25$ & $0.2$ & $0.493528-0.290186 i$ & $0.493528-0.290186 i$ & $0\%$\\
$-0.25$ & $0.3$ & $0.501156-0.279184 i$ & $0.501156-0.279183 i$ & $0.00012\%$\\
$-0.25$ & $0.4$ & $0.511491-0.263396 i$ & $0.511491-0.263395 i$ & $0.00003\%$\\
$-0.25$ & $0.5$ & $0.524029-0.242503 i$ & $0.524028-0.242501 i$ & $0.00032\%$\\
$0$ & $0$ & $0.463851-0.295604 i$ & $0.463851-0.295604 i$ & $0\%$\\
$0$ & $0.1$ & $0.465470-0.293263 i$ & $0.465470-0.293263 i$ & $0\%$\\
$0$ & $0.2$ & $0.470286-0.286177 i$ & $0.470286-0.286177 i$ & $0\%$\\
$0$ & $0.3$ & $0.478157-0.274149 i$ & $0.478157-0.274148 i$ & $0.00014\%$\\
$0$ & $0.4$ & $0.488770-0.256868 i$ & $0.488769-0.256868 i$ & $0.00007\%$\\
$0$ & $0.5$ & $0.501520-0.233988 i$ & $0.501520-0.233987 i$ & $0.00011\%$\\
$0.25$ & $0$ & $0.444443-0.291713 i$ & $0.444443-0.291713 i$ & $0\%$\\
$0.25$ & $0.1$ & $0.446113-0.289200 i$ & $0.446113-0.289200 i$ & $0\%$\\
$0.25$ & $0.2$ & $0.451077-0.281588 i$ & $0.451077-0.281588 i$ & $0.00002\%$\\
$0.25$ & $0.3$ & $0.459169-0.268654 i$ & $0.459169-0.268653 i$ & $0.00012\%$\\
$0.25$ & $0.4$ & $0.470028-0.250047 i$ & $0.470027-0.250046 i$ & $0.00017\%$\\
$0.25$ & $0.5$ & $0.482941-0.225393 i$ & $0.482948-0.225400 i$ & $0.00172\%$\\
$0.5$ & $0$ & $0.427941-0.287605 i$ & $0.427941-0.287605 i$ & $0\%$\\
$0.5$ & $0.1$ & $0.429658-0.284937 i$ & $0.429658-0.284937 i$ & $0\%$\\
$0.5$ & $0.2$ & $0.434758-0.276850 i$ & $0.434758-0.276850 i$ & $0.\times 10^{\text{-4}}\%$\\
$0.5$ & $0.3$ & $0.443054-0.263095 i$ & $0.443054-0.263094 i$ & $0.00021\%$\\
$0.5$ & $0.4$ & $0.454131-0.243279 i$ & $0.454130-0.243278 i$ & $0.00038\%$\\
$0.5$ & $0.5$ & $0.467162-0.217002 i$ & $0.467197-0.217028 i$ & $0.00848\%$\\
$0.75$ & $0$ & $0.413623-0.283485 i$ & $0.413623-0.283485 i$ & $0\%$\\
$0.75$ & $0.1$ & $0.415384-0.280675 i$ & $0.415384-0.280675 i$ & $0\%$\\
$0.75$ & $0.2$ & $0.420611-0.272152 i$ & $0.420611-0.272152 i$ & $0.0001\%$\\
$0.75$ & $0.3$ & $0.429096-0.257638 i$ & $0.429095-0.257637 i$ & $0.0003\%$\\
$0.75$ & $0.4$ & $0.440366-0.236702 i$ & $0.440363-0.236699 i$ & $0.00067\%$\\
$0.75$ & $0.5$ & $0.453470-0.208914 i$ & $0.453601-0.208858 i$ & $0.0287\%$\\
\hline
\end{tabular}%
}}
\end{minipage}
\caption{Representative scalar quasinormal frequencies of the Reissner--Nordstr\"om black hole with $M=1$, computed with the Pad\'e-improved 16th- and 14th-order WKB formulas. Panel (a) shows the $\ell=2$, $n=0$ mode, while panel (b) shows the $\ell=2$, $n=1$ mode. Only representative rows are printed; intermediate rows with weak frequency variation remain commented in the source. The last column gives the relative percent difference between the two WKB estimates.}
\label{tab:qnm-wkb-l2}
\end{table*}

The data in Tables~\ref{tab:qnm-wkb} and \ref{tab:qnm-wkb-l2}, together with Figs.~\ref{fig:damping-mu} and \ref{fig:td-profile}, also indicate the expected accuracy of the WKB results. The internal error estimate supplied by the difference between the 16th- and 14th-order Pad\'e-WKB values depends strongly on the multipole number and on the proximity to the edge of the barrier regime. For the fundamental $\ell=0$ mode the disagreement is already at the percent level for a substantial part of the mass grid and can rise to several percent, reaching about $16\%$ at $(\gamma,\mu)=(0.75,0.18)$. This means that the $\ell=0$ entries, especially near the largest admissible $\mu$, should be regarded as qualitative. For the fundamental $\ell=1$ mode the situation is much better: over most of the grid the two WKB orders differ only at the level of $10^{-3}$--$10^{-1}\%$, and even the largest discrepancy remains below $1\%$ (about $0.88\%$ at $(\gamma,\mu)=(0.75,0.42)$). The eikonal sector is more stable still: for $(\ell,n)=(2,0)$ the WKB14 and WKB16 values coincide to the shown digits throughout the whole table, while for $(\ell,n)=(2,1)$ the difference stays below $3\times10^{-2}\%$ and reaches its maximum only at the endpoint $(\gamma,\mu)=(0.75,0.5)$. The independent time-domain check in Fig.~\ref{fig:td-profile} supports this accuracy estimate, since for $M=1$, $\ell=1$, $\mu=0.1$, and $\gamma=-1$ the Prony and WKB frequencies differ by only $0.0038\%$. Therefore the $\ell\geq1$ data in the barrier region can be considered quantitatively reliable, whereas the $\ell=0$ sector close to the disappearance of the barrier should be treated with caution.

The tables also show a clear and physically consistent dependence on the tidal-charge parameter $\gamma$. For all four mode families the real part $\re{\omega}$ decreases as $\gamma$ increases, so more negative $\gamma$ corresponds to faster oscillations, whereas positive $\gamma$ shifts the spectrum to lower frequencies. The damping rate is more subtle at very small masses, where $\omega_I$ may have a shallow maximum for intermediate negative $\gamma$, but for moderate and large $\mu$ the dominant trend is the one displayed in Fig.~\ref{fig:damping-mu}: increasing $\gamma$ lowers $\omega_I=-\im\omega$ and drives the modes closer to quasi-resonance. This is quantified by Table~\ref{tab:mu-critical}, where for the fundamental family $(\ell,n)=(1,0)$ the extrapolated threshold decreases from $\mu_c\simeq0.697$ at $\gamma=-1$ to $\mu_c\simeq0.465$ at $\gamma=0.75$. The same tendency is visible in Tables~\ref{tab:qnm-wkb} and \ref{tab:qnm-wkb-l2}; for example, at $\mu=0.4$ the fundamental $\ell=1$ mode changes from $0.426243-0.067843 i$ at $\gamma=-1$ to $0.347456-0.032681 i$ at $\gamma=0.75$, and the $(\ell,n)=(2,1)$ mode changes from $0.631538-0.250810 i$ to $0.440366-0.236702 i$. Thus positive tidal charge makes the ringing slower and less damped at fixed mass, while more negative $\gamma$ keeps the oscillation frequency higher and postpones the onset of the long-lived regime to larger scalar-field masses.

\section{Grey-body factors and absorption cross sections}\label{sec:gbf}

The same effective potential that governs the quasinormal ringing also determines the probability that an incoming wave packet is transmitted through the barrier and reaches the horizon. For the scattering problem one keeps the frequency real and considers a wave incident from spatial infinity. To distinguish this real scattering frequency from the generally complex quasinormal frequency $\omega$, we denote it by $\Omega$. Instead of the quasinormal boundary conditions (\ref{boundaryconditions}), one imposes
\begin{equation}\label{gbf-bc-hor}
\Psi\sim \mathcal{T}_\ell(\Omega)e^{-\imo\Omega r_*}, \qquad r_*\to-\infty,
\end{equation}
\begin{equation}\label{gbf-bc-inf}
\Psi\sim e^{-\imo\chi r_*}+\mathcal{R}_\ell(\Omega)e^{+\imo\chi r_*}, \qquad r_*\to+\infty,
\end{equation}
with $\chi=\sqrt{\Omega^2-\mu^2}$ and $\Omega>\mu$ so that the asymptotic solution is oscillatory. Here $\mathcal{R}_\ell$ and $\mathcal{T}_\ell$ are the reflection and transmission amplitudes. Their squared moduli give the reflection probability and the grey-body factor,
\begin{equation}
\Gamma_\ell(\Omega)\equiv |\mathcal{T}_\ell(\Omega)|^2, \qquad |\mathcal{R}_\ell(\Omega)|^2+\Gamma_\ell(\Omega)=1.
\end{equation}
Physically, $\Gamma_\ell$ measures the fraction of the incident flux that penetrates the curvature barrier; equivalently, in the Hawking-emission problem it quantifies the deviation from a perfect blackbody spectrum caused by backscattering off the spacetime geometry.
Here we apply the standard 6th order WKB method for finding grey-body factors which has been widely used and discussed in the literature \cite{Konoplya:2023moy, Konoplya:2023ppx, Konoplya:2021ube, Konoplya:2019ppy, Lutfuoglu:2025blw,  Lutfuoglu:2025ohb, Lutfuoglu:2025ljm,
Lutfuoglu:2025hjy,  Malik:2025czt, Malik:2025qnr,  
Dubinsky:2026wcv, Dubinsky:2025wns, Dubinsky:2025ypj, Dubinsky:2025nxv, Dubinsky:2024nzo, Skvortsova:2024msa}.

In addition, for calculations of grey-body factors we will apply the correspondence between quasinormal modes and grey-body factors.
A particularly useful feature of this correspondence  \cite{Konoplya:2024lir,Konoplya:2024vuj} is that it recasts the resonant data encoded in quasinormal frequencies into a prediction for real-frequency transmission. As long as the effective potential is represented by a single smooth barrier, the same structure near its maximum governs both the WKB spectrum and the grey-body factor. Even though the RN-like background studied here is spherically symmetric, the idea itself is broader: it was originally formulated for asymptotically flat spherical black holes and later extended to rotating, separable geometries in the nonsuperradiant regime \cite{Konoplya:2024vuj}. If we denote the fundamental mode and first overtone by $\omega_0\equiv\omega_{\ell 0}$ and $\omega_1\equiv\omega_{\ell 1}$, then the leading eikonal relation takes the form
\begin{equation}\label{eq:gbf_qnm_eikonal}
i\K_{\ell}(\omega)=\frac{\omega^2-\re{\omega_0}^2}{4\,\re{\omega_0}\,\im{\omega_0}}+\Order{\ell^{-1}},
\end{equation}
whereas the higher-accuracy expression analogous to Eq.~(23) of Ref.~\cite{Konoplya:2024vuj} is
\begin{widetext}
\begin{equation}
\begin{aligned}
i\K_{\ell}(\omega)&=\frac{\omega^2-\re{\omega_0}^2}{4\,\re{\omega_0}\,\im{\omega_0}}
\left(1+\frac{\left(\re{\omega_0}-\re{\omega_1}\right)^2}{32\,\im{\omega_0}^2}
-\frac{3\,\im{\omega_0}-\im{\omega_1}}{24\,\im{\omega_0}}\right)\\
&\quad -\frac{\left(\omega^2-\re{\omega_0}^2\right)^2}{16\,\re{\omega_0}^3\,\im{\omega_0}}
\left(1+\frac{\re{\omega_0}\left(\re{\omega_0}-\re{\omega_1}\right)}{4\,\im{\omega_0}^2}\right)\\
&\quad +\frac{\left(\omega^2-\re{\omega_0}^2\right)^3}{32\,\re{\omega_0}^5\,\im{\omega_0}}\times
\left(1+\frac{\re{\omega_0}\left(\re{\omega_0}-\re{\omega_1}\right)}{4\,\im{\omega_0}^2}
+\frac{\re{\omega_0}^2\left(\re{\omega_0}-\re{\omega_1}\right)^2}{16\,\im{\omega_0}^4}\right.\\
&\qquad \left.-\frac{3\,\im{\omega_0}-\im{\omega_1}}{12\,\im{\omega_0}}\right)+\Order{\ell^{-3}}.
\end{aligned}
\end{equation}
\end{widetext}
This refined mapping has already been examined in a variety of black-hole and regular-spacetime models and, in many cases, remains quantitatively successful even when the multipole number is not especially large \cite{Konoplya:2010vz,Malik:2024wvs,
Dubinsky:2024vbn,Lutfuoglu:2025eik,Lutfuoglu:2025kqp,
Lutfuoglu:2025mqa,Lutfuoglu:2025ldc, Bolokhov:2026eqf,  Bolokhov:2024otn,Malik:2025dxn,Malik:2024cgb}. Nevertheless, its validity is still bounded by the usual WKB assumptions: once the potential no longer looks like a single barrier---for instance, when a double-well profile appears \cite{Konoplya:2025hgp}---or when higher-curvature corrections strongly distort the centrifugal sector and trigger catastrophic instabilities \cite{Konoplya:2017zwo,Takahashi:2010gz,Dotti:2004sh,Konoplya:2017lhs,Konoplya:2017ymp}, one should not expect the grey-body factor reconstructed from quasinormal frequencies to remain trustworthy.

Before turning to absorption cross sections, it is useful to compare the two semiclassical prescriptions for the grey-body factor. Figure~\ref{fig:gbf-methods} shows that for $M=1$ and $\ell=2$ the standard sixth-order WKB result and the estimate obtained via the correspondence with quasinormal modes are almost indistinguishable for all three representative values of $\gamma$. At the same time, increasing $\gamma$ shifts the rise of $\Gamma_\ell(\Omega)$ to lower frequencies, while the difference between the two methods remains small and is concentrated in the transition region where the grey-body factor grows most rapidly.

\begin{widetext}
\refstepcounter{figure}\label{fig:gbf-methods}
\begin{center}
\begin{minipage}{0.48\textwidth}
\centering
\textbf{(a)}\\[-0.4em]
\includegraphics[width=\linewidth]{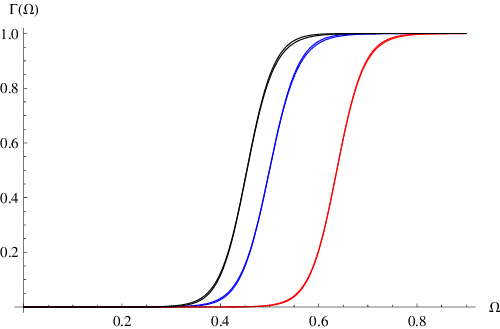}
\end{minipage}\hfill
\begin{minipage}{0.48\textwidth}
\centering
\textbf{(b)}\\[-0.4em]
\includegraphics[width=\linewidth]{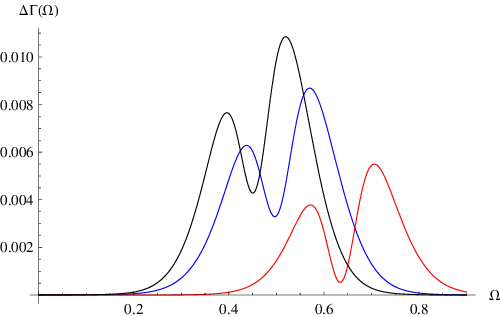}
\end{minipage}\\[0.5em]
\parbox{0.96\textwidth}{\small \textbf{FIG.~\thefigure.} Grey-body factors and the discrepancy between two semiclassical evaluations for the massive scalar field with $M=1$ and $\ell=2$. Panel (a) compares the grey-body factor $\Gamma_\ell(\Omega)$ obtained with the standard sixth-order WKB method and via the correspondence with quasinormal modes. Panel (b) shows the corresponding difference $\Delta \Gamma_\ell(\Omega)$ between the two methods. In both panels the blue, red, and black curves correspond to $\gamma=0$, $\gamma=-1$, and $\gamma=0.75$, respectively.}
\end{center}
\end{widetext}

A complementary comparison is shown in Fig.~\ref{fig:gbf-methods-mu}, where the tidal-charge parameter is fixed at $\gamma=-1$ and the scalar-field mass is varied. For $M=1$ and $\ell=2$, increasing $\mu$ shifts the rise of $\Gamma_\ell(\Omega)$ to higher frequencies, while the discrepancy between the sixth-order WKB result and the estimate based on the correspondence with quasinormal modes remains localized in the transition region and becomes more pronounced for larger $\mu$.

\begin{widetext}
\refstepcounter{figure}\label{fig:gbf-methods-mu}
\begin{center}
\begin{minipage}{0.48\textwidth}
\centering
\textbf{(a)}\\[-0.4em]
\includegraphics[width=\linewidth]{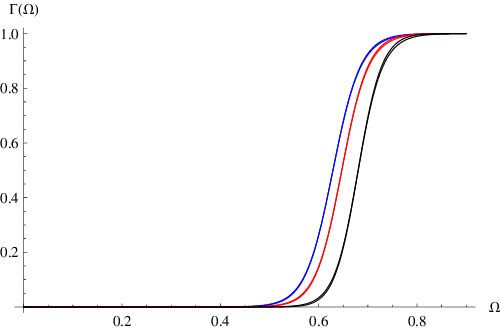}
\end{minipage}\hfill
\begin{minipage}{0.48\textwidth}
\centering
\textbf{(b)}\\[-0.4em]
\includegraphics[width=\linewidth]{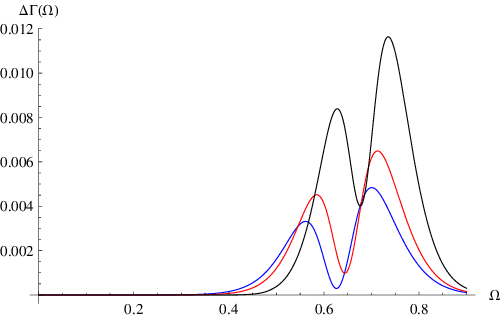}
\end{minipage}\\[0.5em]
\parbox{0.96\textwidth}{\small \textbf{FIG.~\thefigure.} Grey-body factors and the discrepancy between two semiclassical evaluations for the massive scalar field with $M=1$, $\ell=2$, and fixed $\gamma=-1$. Panel (a) compares the grey-body factor $\Gamma_\ell(\Omega)$ obtained with the standard sixth-order WKB method and via the correspondence with quasinormal modes. Panel (b) shows the corresponding difference $\Delta \Gamma_\ell(\Omega)$ between the two methods. In both panels the blue, red, and black curves correspond to $\mu=0$, $\mu=0.3$, and $\mu=0.5$, respectively.}
\end{center}
\end{widetext}

The partial absorption cross section follows from the partial-wave decomposition,
\begin{equation}
\sigma_\ell(\Omega)=\frac{\pi(2\ell+1)}{\chi^2}\Gamma_\ell(\Omega), \qquad \sigma_{\rm abs}(\Omega)=\sum_{\ell=0}^{\infty}\sigma_\ell(\Omega).
\end{equation}
For a massive field the momentum at infinity is $\chi$ rather than $\Omega$, so the threshold $\Omega=\mu$ plays a special role: below it the asymptotic solution becomes evanescent and the standard scattering interpretation is lost.

Figure~\ref{fig:acs-gamma} shows the corresponding partial and total absorption cross sections for a representative fixed mass $\mu=0.2$ and three values of the tidal-charge parameter. In each panel the black curve represents the total absorption cross section, while the colored curves show the partial contributions from individual multipoles.

\begin{widetext}
\refstepcounter{figure}\label{fig:acs-gamma}
\begin{center}
\begin{minipage}{0.32\textwidth}
\centering
\textbf{(a)}\\[-0.4em]
\includegraphics[width=\linewidth]{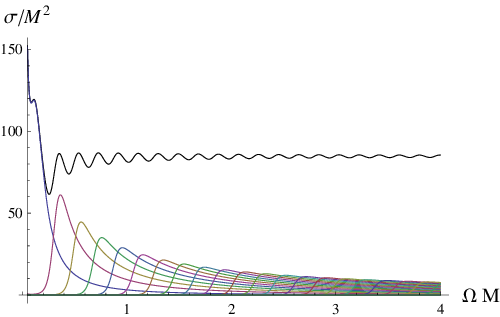}
\end{minipage}\hfill
\begin{minipage}{0.32\textwidth}
\centering
\textbf{(b)}\\[-0.4em]
\includegraphics[width=\linewidth]{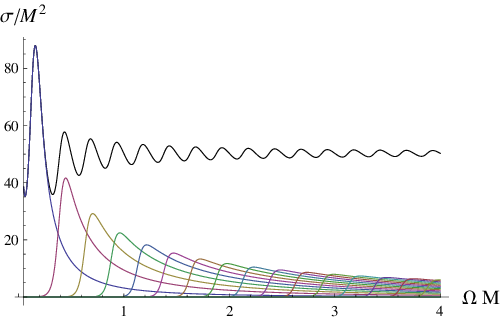}
\end{minipage}\hfill
\begin{minipage}{0.32\textwidth}
\centering
\textbf{(c)}\\[-0.4em]
\includegraphics[width=\linewidth]{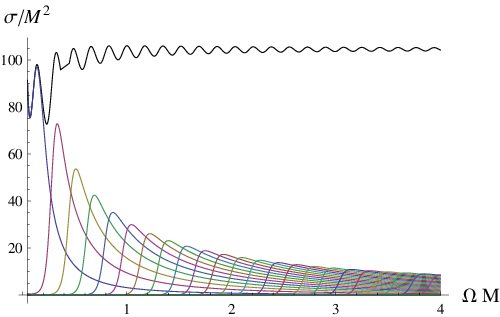}
\end{minipage}\\[0.5em]
\parbox{0.98\textwidth}{\small \textbf{FIG.~\thefigure.} Partial and total absorption cross sections for the massive scalar field with $M=1$ and $\mu=0.2$. In each panel the black curve shows the total absorption cross section $\sigma_{\rm abs}$, while the colored curves show the partial absorption cross sections $\sigma_\ell$ for successive multipoles. Panels (a), (b), and (c) correspond to $\gamma=0$, $\gamma=-1$, and $\gamma=0.75$, respectively.}
\end{center}
\end{widetext}

A second comparison is shown in Fig.~\ref{fig:acs-mu}, where the tidal-charge parameter is fixed at $\gamma=-1$ and the scalar-field mass is varied. In both panels $M=1$, the black curve gives the total absorption cross section, and the colored curves show the partial contributions from individual multipoles.

\begin{widetext}
\refstepcounter{figure}\label{fig:acs-mu}
\begin{center}
\begin{minipage}{0.48\textwidth}
\centering
\textbf{(a)}\\[-0.4em]
\includegraphics[width=\linewidth]{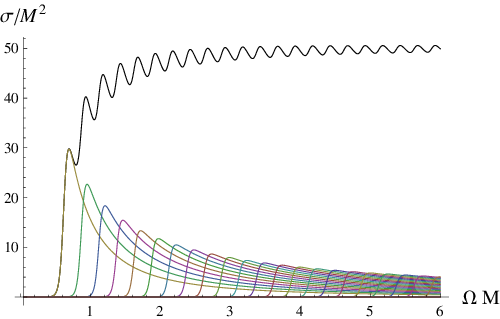}
\end{minipage}\hfill
\begin{minipage}{0.48\textwidth}
\centering
\textbf{(b)}\\[-0.4em]
\includegraphics[width=\linewidth]{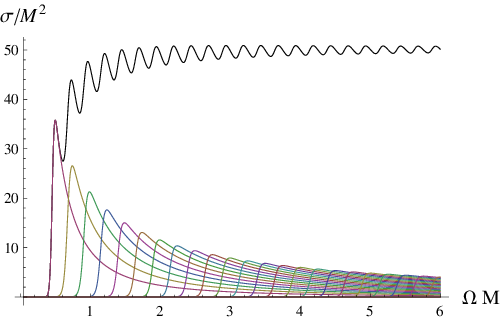}
\end{minipage}\\[0.5em]
\parbox{0.96\textwidth}{\small \textbf{FIG.~\thefigure.} Partial and total absorption cross sections for the massive scalar field with $M=1$ and fixed $\gamma=-1$. In each panel the black curve shows the total absorption cross section $\sigma_{\rm abs}$, while the colored curves show the partial absorption cross sections $\sigma_\ell$ for successive multipoles. Panels (a) and (b) correspond to $\mu=0$ and $\mu=0.5$, respectively.}
\end{center}
\end{widetext}

Within the WKB approach one uses the same expansion about the maximum of the potential as in Sec.~\ref{sec:WKB}, but now $\Omega$ is real and the quantity $\K$ in Eq.~(\ref{WKBformula-spherical}) is treated as a complex parameter rather than being fixed to $n+1/2$ \cite{Konoplya:2019hlu,Iyer:1986np,Konoplya:2003ii,Matyjasek:2017psv}. For the grey-body factors we truncate this expansion at sixth order, that is, we use the standard sixth-order WKB method rather than the Pad\'e-improved quasinormal-mode scheme. After solving Eq.~(\ref{WKBformula-spherical}) for $\K(\Omega)$, the reflection amplitude is approximated by
\begin{equation}\label{Rgbf}
\mathcal{R}_\ell(\Omega)=\left(1+e^{-2\pi \imo \K(\Omega)}\right)^{-1/2},
\end{equation}
and the grey-body factor is obtained from
\begin{equation}\label{Tgbf}
\Gamma_\ell(\Omega)=|\mathcal{T}_\ell(\Omega)|^2=1-|\mathcal{R}_\ell(\Omega)|^2.
\end{equation}
In this way the grey-body factors and absorption cross sections are computed with the standard sixth-order WKB approximation, while Pad\'e improvement is used only for the quasinormal-mode calculations above. This treatment is applied only when the effective potential remains a single smooth barrier.

The behavior of the grey-body factors in Figs.~\ref{fig:gbf-methods} and \ref{fig:gbf-methods-mu} follows directly from the barrier trends in Figs.~\ref{fig:potential-mu} and \ref{fig:potential-params}. At fixed $\ell=2$ and $\mu=0.2$, increasing $\gamma$ lowers the effective-potential peak, so transmission becomes easier and $\Gamma_\ell(\Omega)$ turns on at smaller frequencies; this is why the $\gamma=0.75$ curve lies to the left of the $\gamma=0$ and $\gamma=-1$ curves. By contrast, increasing the field mass raises the asymptotic tail and, in the representative $\gamma=-1$, $\ell=2$ case, also raises the barrier peak, so the wave must tunnel through a more opaque potential and the transition to $\Gamma_\ell\simeq1$ is shifted to larger $\Omega$.

The role of the multipole number is even more direct, because the centrifugal term $\ell(\ell+1)/r^2$ strongly increases the height of the barrier. Higher-$\ell$ partial waves therefore have smaller grey-body factors at a fixed frequency and contribute appreciably to the absorption only at larger $\Omega$. This pattern is visible in Figs.~\ref{fig:acs-gamma} and \ref{fig:acs-mu}: the low-frequency absorption is dominated by the lowest partial waves, while higher multipoles appear as a sequence of peaks shifted toward larger frequencies. Since $\sigma_\ell\propto \Gamma_\ell/\chi^2$, once a partial wave begins to transmit it can produce a pronounced peak near threshold. Lower barriers, as for larger positive $\gamma$, make these channels open earlier and increase the total absorption cross section, whereas heavier fields redistribute the absorption toward higher frequencies.

Figures~\ref{fig:gbf-methods}(a) and \ref{fig:gbf-methods-mu}(a) also show that the correspondence between quasinormal modes and grey-body factors is remarkably accurate in the barrier regime. The two determinations of $\Gamma_\ell$ are practically indistinguishable in the main plots, and the absolute difference $\Delta \Gamma_\ell$ shown in panels (b) remains at the level of $10^{-2}$ or smaller in all displayed cases, with only a modest growth for larger $\mu$. The mismatch is concentrated in the narrow interval where the transmission rises most steeply, while in the opaque regime $\Gamma_\ell\approx0$ and in the transparent regime $\Gamma_\ell\approx1$ the agreement is essentially exact at the scale of the figures.

\section{Conclusions}
We have studied massive scalar perturbations of the Reissner--Nordstr\"om-like brane-world black hole with metric function $f(r)=1-2M/r-\gamma/r^2$ by combining Pad\'e-improved WKB calculations with time-domain integration and, for scattering, the standard sixth-order WKB method. The analysis begins with the exact scalar-field master equation and with a characterization of the effective potential in the $(\gamma,\mu,\ell)$ parameter space. The key geometric result is that the WKB treatment is meaningful only while the potential retains a single barrier. Increasing the scalar-field mass lifts the asymptotic tail and can eventually destroy this barrier, whereas increasing the multipole number strengthens the centrifugal barrier. For the representative configurations relevant to the scattering problem, more negative $\gamma$ raises the barrier and positive $\gamma$ lowers it. In the Schwarzschild $\ell=0$ limit, for example, the barrier disappears already at $\mu=0.25$, which provides a useful benchmark for the onset of the non-WKB regime.

The quasinormal-mode analysis shows that the tidal-charge parameter and the scalar-field mass affect the spectrum in a systematic way. For all families studied, the real part of the frequency decreases as $\gamma$ increases, so more negative $\gamma$ corresponds to faster oscillations, while positive $\gamma$ shifts the spectrum to lower frequencies. At the same time, increasing $\mu$ drives the damping rate toward zero and produces the expected approach to quasi-resonance. For the fundamental $(\ell,n)=(1,0)$ family, quadratic fits to the WKB16 data yield extrapolated quasi-resonant thresholds in the range $\mu_c\simeq0.465$--$0.697$, with $\mu_c$ decreasing monotonically from $\gamma=-1$ to $\gamma=0.75$. Thus positive tidal charge brings the system closer to the long-lived regime at smaller scalar-field masses, whereas more negative $\gamma$ postpones this regime to larger $\mu$.

The accuracy analysis shows that the reliability of the WKB approximation depends strongly on the multipole number and on the distance from the boundary of the barrier regime. The fundamental $\ell=0$ sector becomes only qualitative near the largest admissible masses, but the $\ell\geq1$ results remain quantitatively robust throughout the displayed barrier region. In particular, for the fundamental $\ell=1$ mode the disagreement between the 14th- and 16th-order Pad\'e-WKB values is typically tiny, and the independent time-domain check for $(\ell,\mu,\gamma)=(1,0.1,-1)$ gives a Prony frequency in agreement with the WKB16 result at the level of $0.0038\%$. The eikonal sector is even more stable, which confirms that the reported long-lived behavior is not a numerical artifact of the approximation scheme.

The scattering results are fully consistent with the behavior of the effective potential and the quasinormal spectrum. The grey-body factors obtained from the standard sixth-order WKB method and from the correspondence with quasinormal modes are almost indistinguishable in the barrier regime; in all displayed cases the difference $\Delta\Gamma_\ell$ remains of order $10^{-2}$ or smaller and is localized in the narrow transition interval where the transmission rises most rapidly. Increasing $\gamma$ lowers the barrier, shifts the rise of $\Gamma_\ell(\Omega)$ to smaller frequencies, and enhances the total absorption cross section. By contrast, increasing $\mu$ or $\ell$ makes the barrier more opaque, delays the onset of transmission, and moves the dominant partial-wave peaks in the absorption cross section to higher frequencies. Overall, the RN-like tidal-charge parameter provides a simple way to control both the onset of quasi-resonant ringing and the efficiency of absorption, thereby linking barrier geometry, quasinormal spectra, and grey-body factors in a single consistent picture for massive fields.


\bibliography{bibliography}
\end{document}